\documentclass[usenatbib]{mnras}
\usepackage[T1]{fontenc}
\usepackage{ae,aecompl}

\usepackage{graphicx}	% Including figure files
\usepackage{amsmath}	% Advanced maths commands
\usepackage{amssymb}	% Extra maths symbols
\begin{document}

\title[Orbital configuration of Janus and Epimetheus]{Primordial migration of co-orbital satellites as a mechanism for the horseshoe orbit of Janus\,-\,Epimetheus}

\author[A.  Rodr\'iguez et al.]{
A.  Rodr\'iguez$^{1}$,
J. A. Correa-Otto$^{2}$,
T. A. Michtchenko$^{3}$
\\
% List of institutions
$^{1}$Observat\'orio do Valongo, Universidade Federal do Rio de Janeiro, Ladeira do Pedro Ant\^onio 43, 20080-090, Rio de Janeiro, Brazil\\
$^{2}$Grupo de Ciencias Planetarias, Dpto. de Geof\'isica y Astronom\'ia, FCEFyN, UNSJ - CONICET, Av. J. I. de la Roza 590 oeste, \\
J5402DCS, Rivadavia, San Juan, Argentina\\
$^{3}$Instituto de Astronomia, Geof\'isica e Ci\^encias Atmosf\'ericas, USP, Rua do Mat\~ao 1226, 05508-090, S\~ao Paulo, Brazil}

\date{\today}

\maketitle

\begin{abstract}

We analyze the orbital motion of two natural satellites initially in co-orbital configuration with a third (guiding) satellite embedded into a circumplanetary gas disc and undergoing tidal interactions with the central planet. By solving the exact equations of motion, including the dissipative effects and the mutual gravitational perturbations, we investigate the configuration of the system soon after the guiding satellite is disrupted when crossing the Roche limit with the central planet during its orbital decay. The application is done for the system composed of Saturn and the two small co-orbital satellites Janus and Epimetheus. We perform a large number of numerical simulations, varying the mass of the hypothetical guiding satellite,  Saturn's tidal quality factor and the initial configuration and the masses of the small satellites. Analyzing the orbital configurations of Janus and Epimetheus soon after the disruption of the guiding satellite, we obtain that the mutual horseshoe co-orbital orbits appear as a natural outcome of the numerical simulations.

\end{abstract}


%%%%%%%%%%%%%%%%%%%%%%%%%%%%%%%%%%%%%%%%%%%%%%%%%%

%%%%%%%%%%%%%%%%% BODY OF PAPER %%%%%%%%%%%%%%%%%%

\section{Introduction}\label{intro}

In the frame of the three-body problem, the co-orbital motion, or 1/1 mean motion resonance (MMR), presents two distinct regimes of libration of the resonant angle, defined by the difference between the mean longitudes of the orbiting bodies, $\Delta\lambda$ (see Murray \& Dermott, 1999). When $\Delta\lambda$ librates around 60$^{\circ}$ or 300$^{\circ}$ ($L_4$ and $L_5$ Lagrangian equilibrium solutions, respectively), we have the tadpole regime of motion.  When $\Delta\lambda$ librates with high amplitude around 180$^{\circ}$, enclosing also $L_4$ and $L_5$ Lagrangian points, the regime is known as horseshoe motion. In the phase space of the 1/1 MMR, there exist two separatrices for the libration regimes, one separates tadpole and horseshoe orbits and the other one separates the horseshoe orbit from the circulation of $\Delta\lambda$ (see Giuppone et al., 2010 and Rodr\'iguez et al., 2013).

The tadpole orbits are frequent in the Solar System and, in particular, in the saturnian system of natural satellites. Tethys has two small co-orbital satellites, namely, Telesto ($L_4$) and Calypso ($L_5$), whereas Dione has two others, Helene ($L_4$) and Polydeuces ($L_5$). Contrarily, the horseshoe orbits are rare: with exception of a few asteroids co-orbital with Earth (De la Fuente Marcos \& De la Fuente Marcos, 2016), only Janus and Epimetheus are found in the mutual horseshoe regime of motion (Dermott \& Murray, 1981); they are small moons of Saturn orbiting the planet at 2.51 Saturn's radii on average, sharing the same orbit and swapping their positions every 4 years (Yoder et al., 1983).  This difference can be understood from evolutionary models, which predict the formation and dynamical evolution of the systems close to the stationary configurations, particularly, around the points $L_4$ and $L_5$ (e.g. Jewitt et al., 2004). The horseshoe orbits are, contrarily, peculiar energetic configurations, whose dynamical evolution under external perturbations is easily transformed in a circulation regime of motion (see Rodr\'iguez et al., 2013).

The subject that we are dealing with in this paper is to identify the mechanism that could originate the Janus-Epimetheus mutual horseshoe regime of motion. To our knowledge, up to now, there are no models providing a convincing explanation for the origin of the orbital motion of these small moons, although some works have recently studied the formation of Janus and Epimetheus. Charnoz et al. (2010) and Crida \& El Moutamid (2017) suggested that Janus and Epimetheus were formed through aggregation from Saturn's rings material, whereas Treffenstadt et al. (2015) discussed whether both satellites might have been formed by a collisional disruption of a parental object and evolved into a mutual co-orbital configuration.

Concerning the formation of the saturnian system, Canup \& Ward (2006) investigated the possibility that multiple generations of Titan-sized satellites have been lost as a consequence of tidal and gas-driven migration in a primordial circumplanetary disc. Moreover, Canup (2010) argued that the mid-sized saturnian moons, up to Tethys, may have formed by the removal mass from a Titan-sized satellite who crossed its Roche limit. In this context, we purpose the hypothesis that Janus and Epimetheus were co-orbital satellites of an ancient large (guiding) satellite, which was lost during its Roche limit crossing. The fact that Tethys and Dione have their own small co-orbital satellites enforces our hypothesis.

Assuming a primordial circumplanetary disc and tidal interactions, as well as mutual gravitational perturbations, we investigate the orbital configuration of Janus and Epimetheus after the tidal disruption of their hypothetical guiding satellite. We analyze whether both minor satellites might end in a mutual horseshoe regime of motion, testing a large number of initial configurations by varying different parameters involved in the model. For sake of completeness, we also consider more and less massive satellites than Janus and Epimetheus, by an order of magnitude, investigating the implications on the final orbital configuration.

This paper is organized as follows: in Sec. \ref{model} we describe the scenario and the hypothesis adopted in the paper. The interactions considered in the model are detailed in Sec. \ref{methods}. Sections \ref{num} and \ref{results} display the set of parameters used and the results of the numerical simulations. Finally, Sec.\ref{conclusions} is devoted to discussions and conclusions.

\section{The scenario adopted}\label{model}

\begin{figure}
\centering
\includegraphics[width=0.35\textwidth]{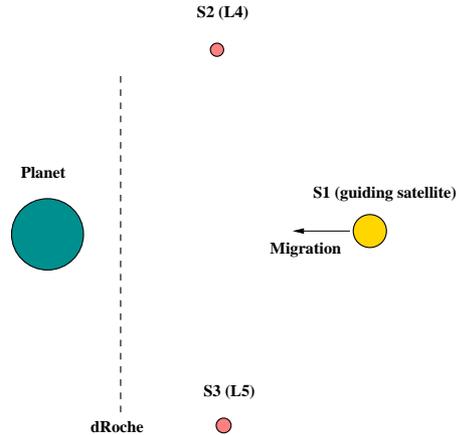}
\caption{{\small Illustration of the model. A main satellite (S1) is orbiting the central planet and have two minor satellites sharing the same orbit in co-orbital motion (S2 and S3). These satellites are initially located around the equilibrium points $L_4$ and $L_5$. S1 migrates towards the planets due to the interaction with a circumplanetary gas disc and tidal forces with the planet, approaching to its Roche limit ($d_{Roche}$) as S2 and S3 migrates together, maintaining the co-orbital configuration.}}
\label{modelo}
\end{figure}

We consider a system composed of a central planet, a main satellite (S1) and two minor satellites (S2 and S3),  in such a way that the masses of S2 and S3 are much smaller than the mass of S1. All satellites are initially in co-orbital configuration. We assume that the small satellites are located near the $L_4$ (S2) and $L_5$ (S3) triangular equilibrium points (see Fig. \ref{modelo}). Following Canup \& Ward (2006) and Canup (2010), we adopt a scenario in which the satellites are embedded into a primordial circumplanetary gas disc. This scenario describes the early migration of regular satellites around giant planets corresponding to the final stages of satellite formation. The tidal interaction of the satellite with the central planet is also taken into account, in such a way that the combined effects of disc and tides result in the satellite migration towards the planet (see next section for details).

We assume that the three satellites have an icy composition similar to the mid-sized moons of Saturn (see Table 1 of Canup 2010). The whole system evolves under mutual gravitational perturbation, tidal and disc interactions. We ignore the gravitational influence of companion planets and satellites, restricting the orbital motion to the coplanar case. For the planet-satellite distances that we will consider here, the central star's gravitational perturbation can also be ignored (see Giuppone and Roig, 2018).

Due to the dissipative forces the satellites undergo inward migration and the Roche limit of S1 ($d_{roche}$) may eventually be crossed, leading to the tidal destruction of the non-differentiated main satellite. We consider that the main satellite is totally disrupted when its Roche limit is attained, while the small satellites continue to evolve under the Keplerian attraction with the central planet and their mutual gravitational perturbation. The aim of this work is to investigate the orbital configuration of the minor satellites when the main migrating satellite crosses its Roche limit.

\section{The model}\label{methods}

\subsection{Gas disc}

In the model of Canup \& Ward (2006) for gas giant satellite formation, the interaction of the the satellites with the gas disc is modeled by a Type I migration. This type of migration arises due to density waves interactions between the satellite and the circumplanetary gas disc, resulting in orbital decay. In this case, for a satellite at a distance $r$ form the central planet, the timescale of inward migration is given by

\begin{equation}
\tau_{disc} = \dfrac{a_i}{\dot{a}_i} \sim \sqrt{\dfrac{m_P^3}{ \mathcal{G} r}} \dfrac{ h^2}{C_a  m_i \sigma_G} \rm{,}
\label{eqdisc1}
\end{equation}
where $a_i$ is the semi-major axis of the satellite, $r$ is the instantaneous distance between the planet and the satellite, $m_P$ and $m_i$ are the masses of the planet and the satellite, respectively, $C_a\sim 2.7 + 1.1q_G$, with $q_G$ being the surface density profile of the disc ($q_G=0.5$, according Canup \& Ward 2006), $\sigma_G$ is the gas surface density of the circumplanetary disc and $\mathcal{G}$ is the gravitational constant. In addition, $h=c_s/\Omega r$ is the vertical height scale or the aspect ratio of the disc, with $\Omega$ being the angular rotation velocity of the local gas and $c_s$ is the sound speed. For cold discs, which are rotationally supported, the sound speed $c_s$ is smaller than the rotation velocity $v_{rot}=\Omega r$, and $h \sim 0.05$. Following Canup (2010), we assume that Saturn's rings are in the process of formation, as a consequence of a generation of tidally disrupted moons, and ignore their interaction with the satellites.

In Eq. (\ref{eqdisc1}), the Type I migration timescale ($\tau_{disc}$) is inversely proportional to the mass of the satellite ($m_i$), hence, smaller satellites migrate slower. In this way, for the masses adopted in this work (see Sec. \ref{num}), the rate of orbital decay of the guiding satellite is larger by a factor of 100 when compared to the corresponding rate for the minor satellites. This factor allows us to ignore the influence of the gas disc on the motion of  the S2 and S3 satellites.

In our model, $\sigma_G$ can be estimated through a quasi steady-state balance between the inflow supply of gas to the disc and the gas removal as the disc viscously spreads into the planet or beyond the outer edge of the disc. In the final stage of the disc evolution, the surface density is given by

\begin{equation}
\sigma_G (r) \sim 10 g/cm^2 \sqrt{\dfrac{1.7 R_P}{r}} \dfrac{0.003}{\alpha} \dfrac{0.05}{h}^{2} \dfrac{F_{in}}{10^{-7} M_{\odot} yr^{-1}}
\label{eqdisc2}
\end{equation}
(Canup, 2010), where $R_p$ is the planetary radius, $\alpha \sim 0.001$ is the differential rotation of the disc with respect to the Keplerian velocity, $h\sim0.05$ is the height scale of the disc and $F_{in}$ is the mass inflow rate from the protoplanetary disc, which, for Saturn, is $F_{in} \sim  8 \times 10^{-8} M_{\odot} yr^{-1}$ (Canup, 2010).

\subsection{Tidal forces}

We use the classical Darwin's formalism to describe the planet-satellite tidal interaction (see Ferraz-Mello et al. 2008, for a review). In this context, the averaged variation of the satellite semi-major axis ($a_i$) raised on the planet by the satellite  due to tides, is given by

\begin{equation}
\dot{a}_i=\frac{3n_ik_{2P}m_iR_P^5}{m_Pa_i^4}\varepsilon_0,
\label{tide-gral}
\end{equation}
where $k_{2P}$ is the second-order Love number and radius of the planet and $n_i$ is the mean orbital motion of the satellite (above equation holds for small eccentricity, see equation (56) of Ferraz-Mello et al. 2008). $\varepsilon_0$ is the tidal phase-lag, which is related to the corresponding tidal frequency $\nu_o=2(\omega_P-n_i)$, with $\omega_P$ being the angular rotation velocity of the planet. Usually, a linear tidal theory is adopted in such a way that $\varepsilon_0=\nu_0\Delta t$, where $\Delta t>0$ is the so-called time lag (Mignard 1979; Hut 1981). Hence, the orbital migration due to tides will result in orbital decay or expansion, depending on the sign of $\nu_0$. For distances from the planet above the synchronous radius ($\omega_P=n_i$), we have $\nu_0<0$ and, thus, $\dot{a}_i<0$, resulting in orbital decay in a timescale given by

\begin{equation}
\tau_{tide}=\frac{a_i}{\dot{a}_i}=\frac{1}{3n_i} \frac{Q_P}{k_{2P}}\frac{m_P}{m_i}\left(\frac{a_i}{R_P}\right)^5,
\label{eqtid}
\end{equation}
where we have replaced $k_{2P}\,\varepsilon_0$ by $k_{2P}/Q_P$, with $Q_P$ widely used as the tidal quality factor, accounting for the dissipation in the interior of the tidally deformed body.

A simple inspection of Eq. (\ref{eqtid}) shows that the timescale of orbital decay due to tides in the planet is inversely proportional to $m_i$, the mass of satellite, hence, the effect is more important for large satellites. For this reason, we consider only the tidal force acting on the main satellite S1, that corresponds to $i=1$ in Eq. (\ref{eqtid}).

It is important to note that the semi-major axis will also decrease due to tides raised by the planet on the satellite, however, this contribution is proportional to the square of the eccentricity of the satellite orbit (see Ferraz-Mello et al. 2008) and thus can be neglected for small eccentricities.

\section{Numerical simulations}\label{num}

Here we choose Saturn as being the central planet, so $m_P=m_{Sat}$, where $m_{Sat}$ is the masss of Saturn. For the minor co-orbital satellites S2 and S3, the masses are $3\times10^{-9}m_{Sat}$ and $1\times10^{-9}m_{Sat}$, respectively, which are close to the masses of Janus and Epimetheus\footnote{https://ssd.jpl.nasa.gov/horizons.cgi}, respectively.

The main satellite S1 is assumed to be homogeneous and have an icy body composition. The Roche limit for an homogeneous (and fluid body) satellite is given by the classical expression $d_{Roche}=2.44R_P(\rho_P/\rho_1)^{1/3}$, where $\rho$ stands for mean density (Roche, 1847). As shown in Leinhardt et al. (2012), an homogeneous satellite is fully destroyed when its Roche limit is attained during  the inward migration.

On the other hand, in the case of  Janus and Epimetheus, the disruption limit is smaller than the classical Roche limit, $d_{Roche}$ (Holsapple and Michel, 2006, 2008; Sharma, 2009). Moreover, according to Porco et al. (2007), there exist the possibility for Janus and Epimetheus to be rubble-pile like objects. Theoretical and numerical works suggest that the Roche limit for a rubble-pile satellite is closer to the planet than for a fluid body of the same mean density (Leinhardt et al., 2012, Zhang et al., 2018). Thus, we assume that S2 and S3 are not disrupted when S1 crosses its Roche limit.

The synchronous radius for the current Saturn's rotation velocity is $d_{sync}=1.9R_{Sat}$, where $R_{Sat}$ is Saturn's radius (60,268 km). In addition, replacing values for the current Saturn and an icy satellite, we have $d_{Roche}=2.15R_{Sat}$. Thus, since $d_{sync}<d_{Roche}$, an icy undifferentiated satellite would not undergo inward migration. However, as claimed in Canup (2010), the size and rotation of the early Saturn were different from the current ones, modifying both the synchronous radius and Roche limit values. Following Canup (2010), the synchronous radius was close to 3 -- 3.5 $R_{Sat}$ and, adopting $R_P=1.5R_{Sat}$ (Canup, 2010), we have $d_{Roche}=2.08R_{Sat}$. Hence, $d_{sync}>d_{Roche}$, allowing for inward migration for $a_1\simeq d_{sync}$.

Regarding the expressions for the forces accounting for disc driven and tidal migration, the reader is referred to Tanaka et al. (2002), Tanaka and Ward (2004) and Mignard (1979).

\subsection{Initial conditions}\label{init}

For the disc, we choose $\alpha=0.001$ and $h=$0.05, thus the surface density of the gas disc is given by

\begin{equation}
\sigma_G (r) \sim 30 g/cm^2 \sqrt{\dfrac{1.7 R_P}{r}}.
\label{eqdisc2}
\end{equation}
Since we consider distances between 2.08 $-$ 2.3 $R_{Sat}$ in our numerical simulations, we have $\sigma_G \sim 25 g/cm^2 $.

The mass of S1 is a free parameter in our model and the dissipation within Saturn is not well constrained, thus, we perform several sets of numerical simulations covering a grid in $m_1$ and $(k_2/Q)_P$. The results are shown on a $30\times20$ grid of initial conditions, allowing us to investigate whether mutual horseshoe co-orbital configurations of S2 and S3 are a possible outcome in the simulations, depending on these parameters.

We vary the mass of S1 (the guiding satellite) in the range (0.1 -- 1)$\times10^{-6}m_{Sat}$. This mass range roughly corresponds to (0.1 -- 1)$\times$ Tethys' mass, one of the icy mid-sized moons of Saturn. We assume a constant density of 1 g/cm$^3$ (around the Tethys' density), for all considered mass values of S1.

Classical works adopt $Q_P>18,000$, based on the assumption that the icy satellites from Mimas to Rhea are primordial in their origin (Goldreich 1965, Peale et al. 1980, Meyer and Wisdom 2007). Recently, evidence for strong tidal dissipation within Saturn was pointed out, equivalent to $(k_2/Q)_P=1.6\times10^{-4}$, based on astrometrical measurements of Saturn's mid-sized icy moons (Lainey et al. 2012, 2017). For Saturn, $k_2=0.34$ (Gavrilov and Zharkov, 1977) and the above value of $Q_P$ is of the order of 2,000. Therefore, in our numerical simulations, we change $(k_2/Q)_P$, adopting values in the range (0.1 -- 1.5)$\times10^{-4}$, thus, varying $Q_P$ within the range of possible values adopted in classical and recent works.

The initial angular configuration of the system is such that S2 and S3 are placed in the exact triangular equilibrium points, namely, $L_4$ and $L_5$, with respect to the principal S1 satellite. We set $\lambda_1=0$, $\lambda_2=60^{\circ}$ and $\lambda_3=300^{\circ}$, thus, $\Delta\lambda=\lambda_3-\lambda_2=240^{\circ}$. We also consider small deviations from $L_4$ and $L_5$ as initial angular positions of S2 and S3 (see Table \ref{tab1}).

In order to not restrict our results to the specific case of Janus and Epimetheus, we change (increasing and decreasing) the masses of the minor satellites by one order of magnitude.

All simulations start with $a_1=a_2=a_3=2.3R_{Sat}$ (close to the Roche limit for S1, at $2.08R_{Sat}$) and $e_1=e_2=e_3=0.01$, where $e_i$ ($i$=1,2,3) are the orbital eccentricities of the satellites. As mentioned above, the early radius of Saturn is $1.5R_{Sat}$ and the initial rotation period is 27.5 h, corresponding to the location of the synchronous radius at $3.5R_{Sat}$ (Canup, 2010).  The integration of the equations of motion is performed using a Bulirsch-Stoer integrator with a precision of 10$^{-13}$.

\section{Results}\label{results}

The results of the numerical simulations show that, after S1 reaches its Roche limit and is disrupted, the regime of motion of the minor satellites is characterized by the behavior of the angle $\Delta \lambda$. The final results of our simulations can be classified in two configurations:
\begin{itemize}
\item Regime of mutual tadpole co-orbital motion: $\Delta \lambda$ oscillates around 300$^\circ$ (60$^\circ$).
\item Regime of mutual horseshoe co-orbital configuration: $\Delta \lambda$ oscillates with large amplitude around 180$^\circ$.
\end{itemize}

 \begin{figure}
\centering
\includegraphics[width=0.49\textwidth]{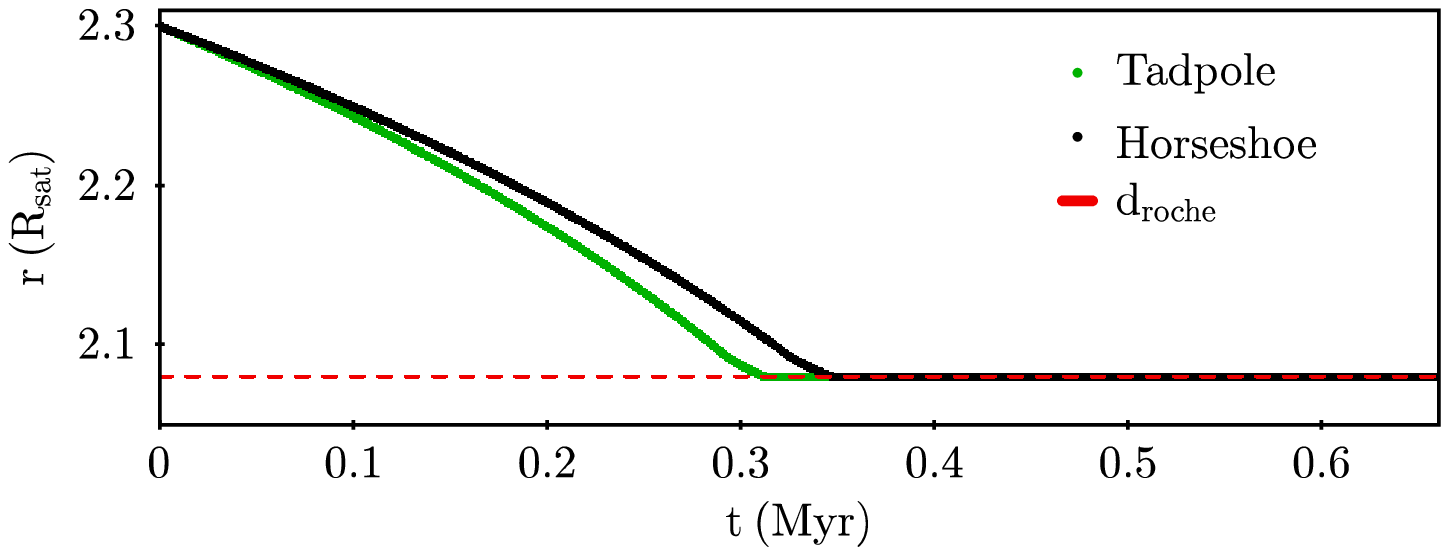}
\includegraphics[width=0.49\textwidth]{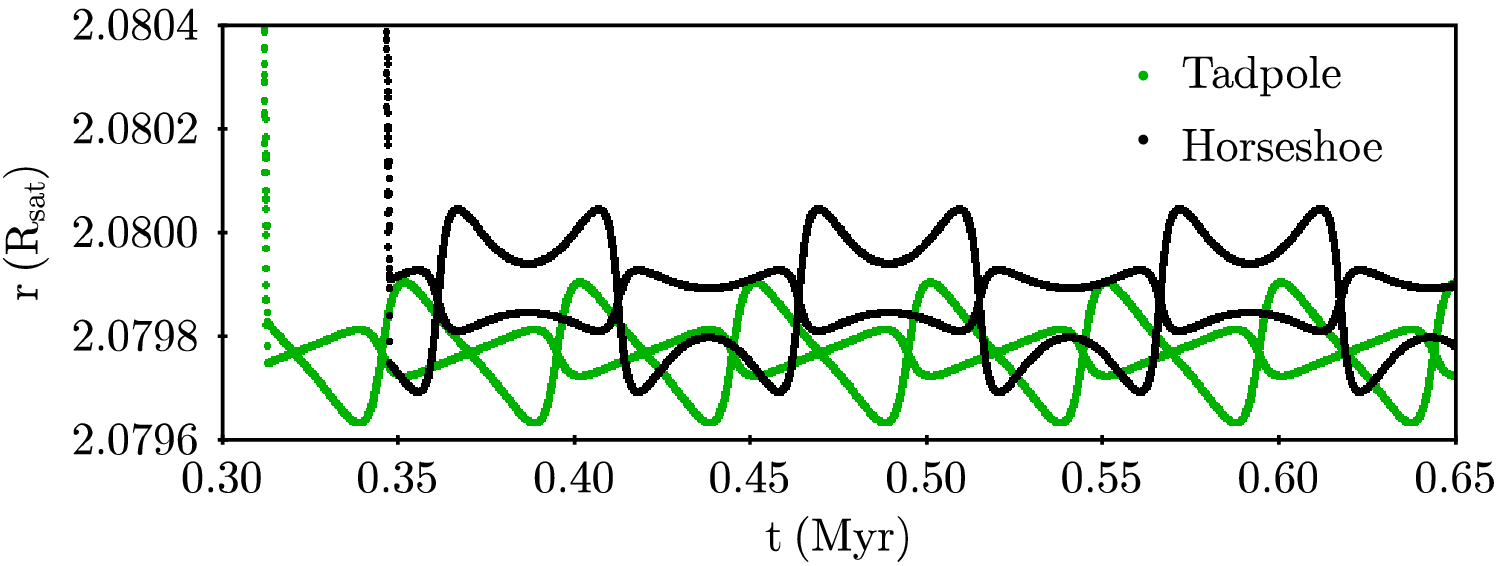}
%\subfigure{\includegraphics[width=0.43\textwidth]{jup3.eps}}
%\includegraphics[width=0.47\textwidth]{ejemplo1.eps}
%\includegraphics [width=0.4\textwidth] {figura0.eps}
\caption{{\small Time variation of the planet-satellite distance for two different initial conditions in the mass of the main satellite (see text for details). The masses of S2 and S3 were taken as the same masses of Janus and Epimetheus. The orbital decay results in the crossing of the Roche limit of S1 ($d_{Roche}$), indicated by the red horizontal line, leading to its tidal disruption. The orbital configurations of the minor satellites after the reaching of $d_{Roche}$ are mutual tadpole (green) and horseshoe (black) regimes of co-orbital motion. The top panel shows the full evolution (including the three satellites), whereas the bottom panel shows a zoom after S1 crosses its Roche limit.}}
\label{figjup}
\end{figure}

\begin{figure}
\centering
\includegraphics[width=0.49\textwidth]{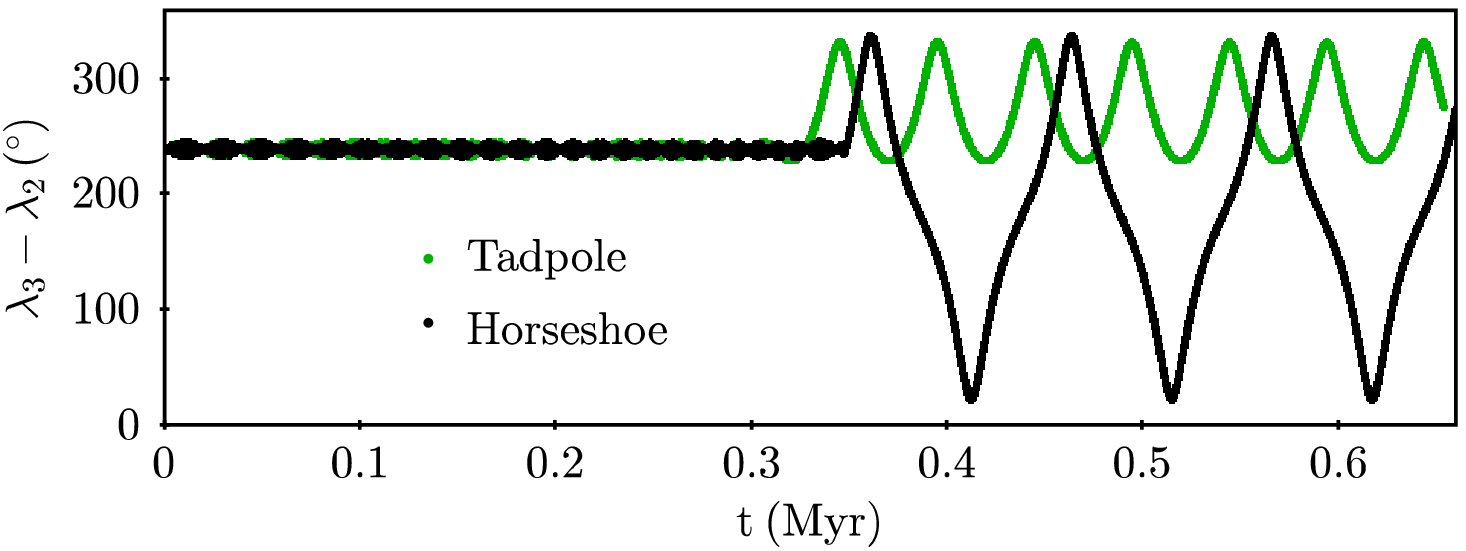}
\includegraphics[width=0.49\textwidth]{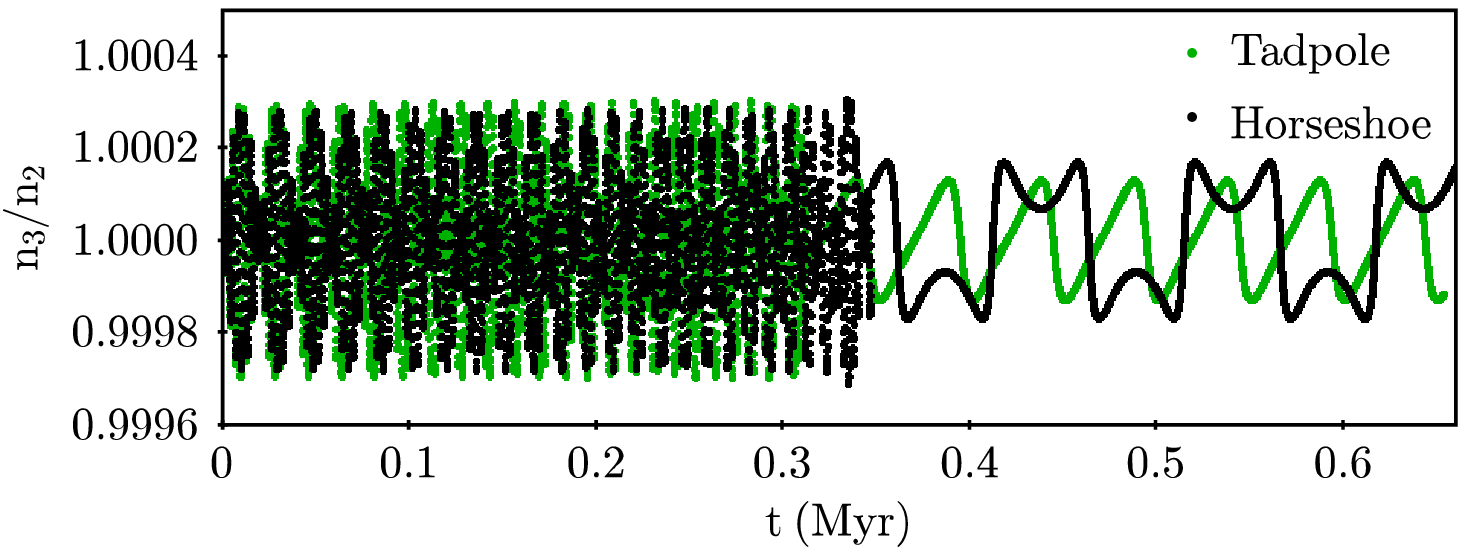}
%\subfigure{\includegraphics[width=0.43\textwidth]{sat3.eps}}
%\includegraphics[width=0.47\textwidth]{ejemplo1.eps}
%\includegraphics [width=0.4\textwidth] {figura0.eps}
\caption{{\small The same as Fig. \ref{figjup} but here showing the time evolution of the resonant angle of the 1/1 MMR (top panel) and mean motion ratio between S2 and S3 (bottom panel). Note as the horseshoe regime is defined by the high amplitude oscillation of $\Delta\lambda$ around 180$^{\circ}$.}}
\label{figsat}
\end{figure}

We show in Figs. \ref{figjup} and \ref{figsat} examples of possibles final configurations considering two cases with the following initial conditions: $a_1$ = 2.3 $R_{Sat}$, $\Delta \lambda=\lambda_3 - \lambda_2 =240^\circ$, $(k_2/Q)_P =$ 10$^{-5}$ and  $\sigma_G \sim 25$ g/cm$^2$, only changing the mass of the main satellite:  $m_1=1\times$ 10$^{-6} m_P$ for the first case (green curves) and $m_1=$ 0.9 $\times$ 10$^{-6} m_P$ for the second case (black curves).

Figures \ref{figjup} and \ref{figsat} show the temporal evolution of semi-major axes, $\Delta \lambda$ and mean motion ratio between the minor satellites corresponding to the two above examples. The orbital decay leads to the crossing of the Roche limit of S1 ($d_{Roche}$), where the main satellite is disrupted at around 0.325 Myr and 0.35 Myr, for each case. The difference in the timescale for achieving the Roche limit is explained by the difference of the main satellite's mass ($m_1$), that affects the migration velocity (see Eq. (\ref{eqtid})). After $d_{Roche}$ is attained, either tadpole (green) or horseshoe (black) co-orbital motions are obtained as final orbital configurations, depending on the mass of S1.

\begin{figure}
\centering
\includegraphics[width=0.43\textwidth]{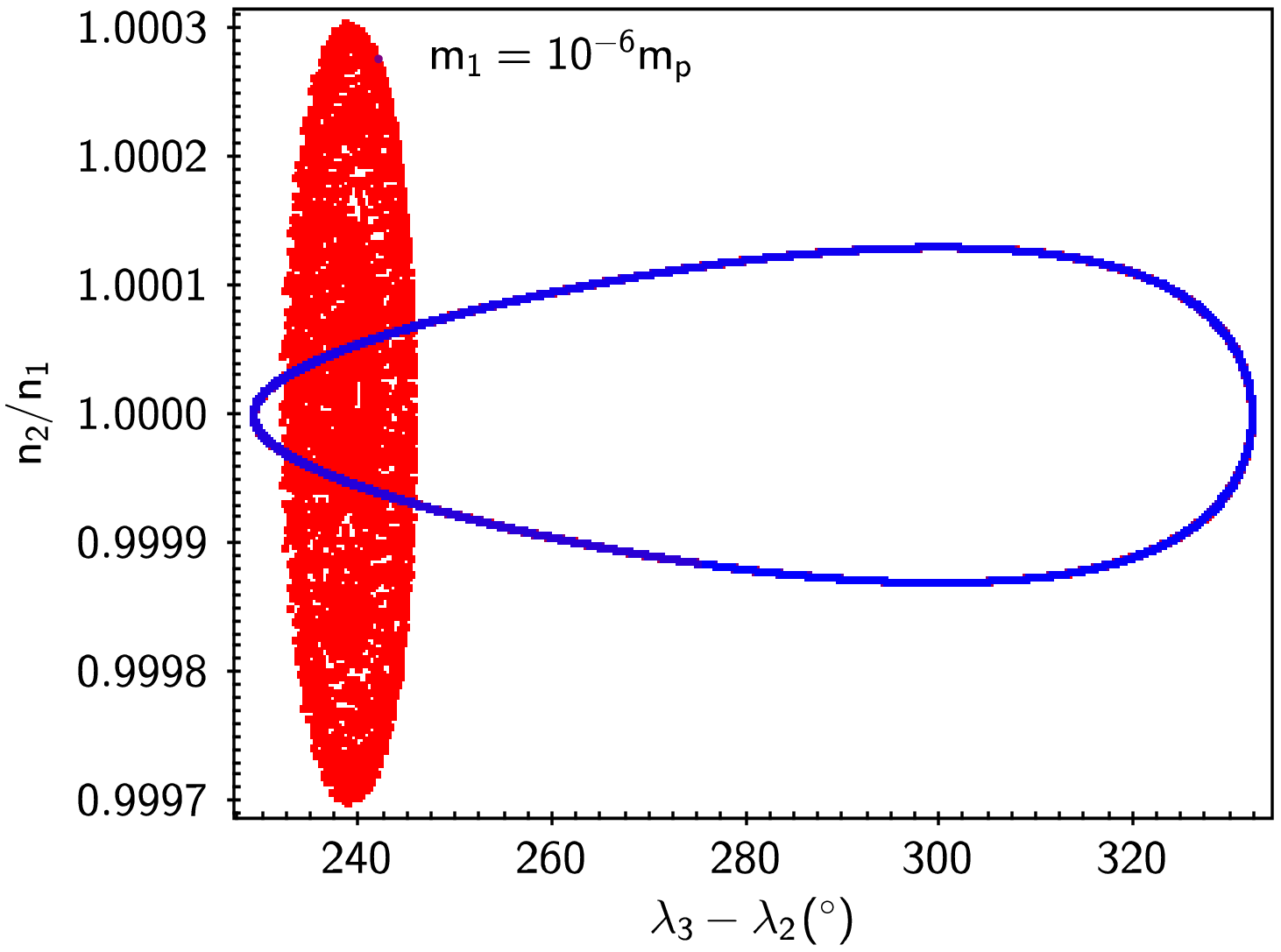}
\includegraphics[width=0.43\textwidth]{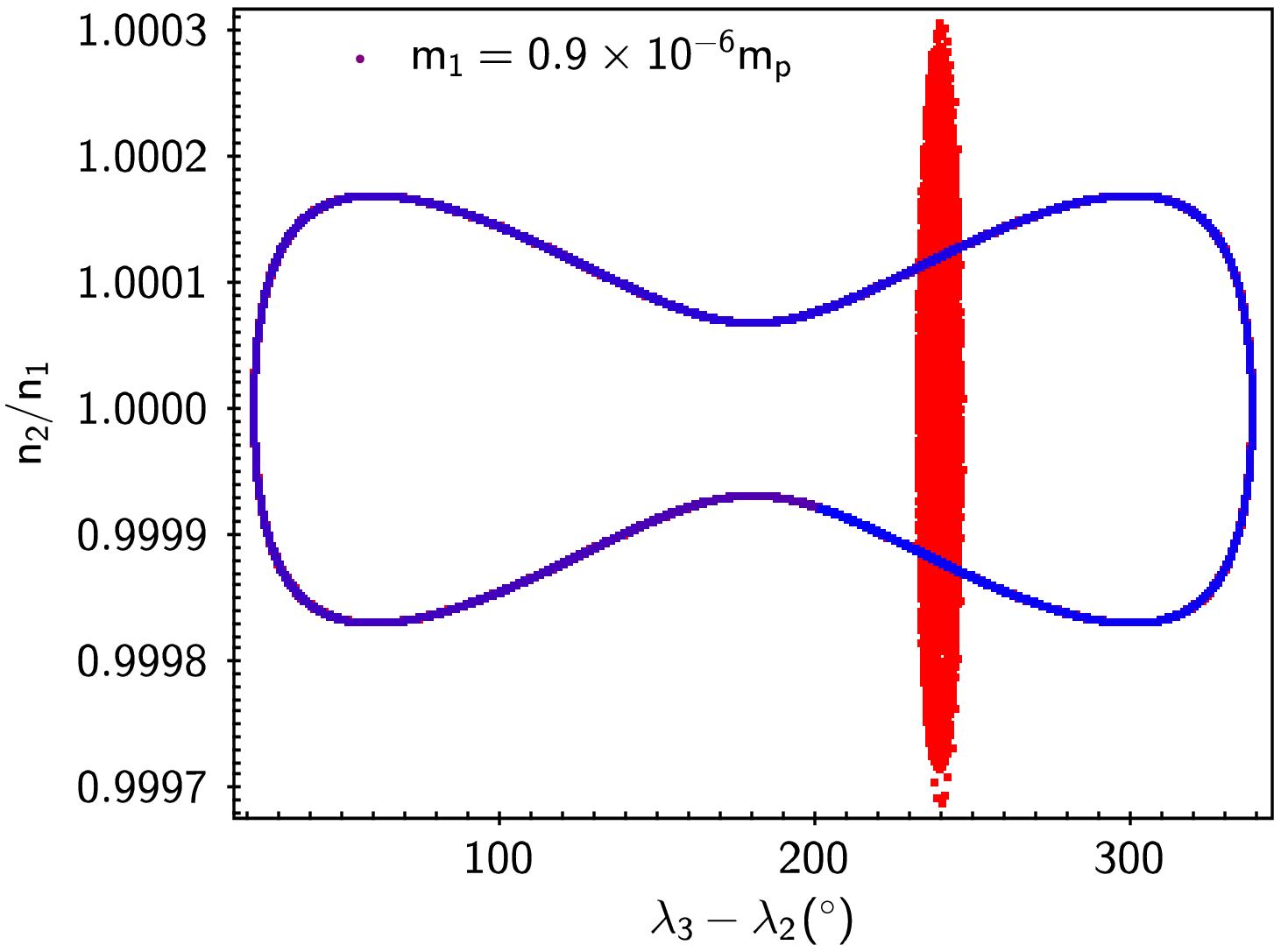}
%\subfigure{\includegraphics[width=0.43\textwidth]{sat3.eps}}
%\includegraphics[width=0.47\textwidth]{ejemplo1.eps}
%\includegraphics [width=0.4\textwidth] {figura0.eps}
\caption{{\small Evolution in the space of resonant angle and mean motion ratio between S2 and S3. Red (blue) points corresponds to the evolution before (after) the Roche limit be attained, for the same initial conditions of Fig. \ref{figjup}. For distances larger than $d_{Roche}$, S2 and S3 evolve in tadpole co-orbital configuration with S1. When $d_{Roche}$ is attained, the tadpole and horseshoe mutual co-orbital configurations between S2 and S3 naturally arise.}}
\label{figsatq}
\end{figure}

Figure \ref{figsatq} shows, for the same examples, the evolution in the phase space of the 1/1 MMR, where we note the oscillation regimes of the angle $\Delta\lambda$ for both types of co-orbital motions. Red (blue) dots corresponds to the motion before (after) the reaching of $d_{Roche}$. It is interesting to note that, until S1 reaches its Roche limit, $\Delta\lambda$ oscillates with small amplitude around 240$^{\circ}$, implying in small oscillations of S2 and S3 around $L_4$ and $L_5$, respectively, in a tadpole configuration with the guiding body. After the Roche limit, the mutual tadpole (top panel) and horseshoe (bottom panel) co-orbital regimes are obtained in this individual numerical simulations. We note that a small change in the mass $m_1$ (from $\times$ 10$^{-6} m_P$ to 0.9 $\times$ 10$^{-6} m_P$) results in two different final orbital configurations of the minor satellites.

\begin{table}
	\centering
	\caption{Initial conditions used in each set of numerical simulations for the masses of the smallest satellites and their angular configurations. Initially, $\lambda_1=0^{\circ}$ for all cases. The mass of the largest satellite and the parameter that controls the orbital decay were changed considering a grid ranging from $0.1\times10^{-6}m_P\leq m_1\leq1\times10^{-6}m_P$ and $0.1\times10^{-4}\leq (k_2/Q)_P\leq1.5\times10^{-4}$. In Simulations 1 to 5, the masses $m_2$ and $m_3$ correspond to the masses of Janus end Epimetheus, respectively. The last column displays the resulting horseshoe mutual co-orbital configurations after the largest satellite is tidally disrupted as it crosses its Roche limit with the central planet.}
	\label{tab1}
	\begin{tabular}{c|cc|cc|cc} % four columns, alignment for each
		\hline \hline
		\\
Simulation    & \multicolumn{2}{|c}{initial} & \multicolumn{2}{|c}{satellite} & \multicolumn{2}{|c}{horseshoe}  \\
               & \multicolumn{2}{|c}{angles} & \multicolumn{2}{|c}{masses} & \multicolumn{2}{|c}{orbits} \\
		\hline		
		\\
		& $\lambda_2$ & $\lambda_3$ & $m_2$ & $m_3$ & \\
		&  ($^\circ$) &  ($^\circ$) &  \multicolumn{2}{|c}({$10^{-9}\,m_P$)} & \% \\
		\hline \\
1 & 60   & 300  &   3 & 1 & 26.0\\
2 & 62   & 300  &   3 & 1 & 37.8\\
3 & 60   & 298  &   3 & 1 & 19.0\\
4 & 62   & 298  &   3 & 1 & 42.1\\
5 & 58   & 298  &   3 & 1 & 6.73\\
6 & 60   & 300  &   0.3 & 0.1 & 65.0\\
7 & 60   & 300  &   30 & 10 & 0.0\\

	    \hline
	\end{tabular}
\end{table}

\begin{figure}
\centering
%\subfigure{\includegraphics[width=0.43\textwidth]{mapa.eps}}%
%\subfigure{\includegraphics[width=0.43\textwidth]{mapa.eps}}
%\subfigure{\includegraphics[width=0.43\textwidth]{fig1d.eps}}
\includegraphics[width=0.47\textwidth]{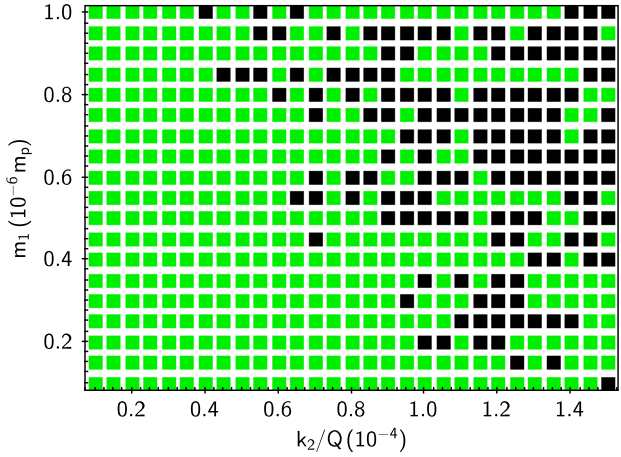}

\caption{{\small Numerical simulations for a grid of initial conditions in $(k_2/Q)_P$ and $m_1$, considering $m_2$ and $m_3$ as roughly being the masses of Janus and Epimetheus. Black (green) squares correspond to initial conditions resulting in horseshoe (tadpole) orbits for the motion of the minor satellites after the main satellite crosses the Roche limit. For all runs we have assumed $e_1 = e_2 = e_3 =$ 0.01 as initial values of eccentricities. In addition, we consider $\lambda_1 =$ 0, $\lambda_2=60^{\circ}$ and $\lambda_3=300^{\circ}$ (S2 and S3 initially located at the $L_4$ and $L_5$ equilibrium points).}}
\label{fig1}
\end{figure}

The results obtained for the final orbital configurations varying $(k_2/Q)_P$ and $m_1$ according to described in Sec. \ref{init}, is shown in Fig. \ref{fig1}. We consider $m_2=3\times10^{-9}m_P$ and $m_3=1\times10^{-9}m_P$ (approximately the masses of Janus and Epimetheus, respectively), whereas $\lambda_1=0^{\circ}$, $\lambda_2=60^{\circ}$ and $\lambda_3=300^{\circ}$, that is, S2 and S3 placed in the exact $L_4$ and $L_5$ equilibrium points with respect to S1, respectively. This case corresponds to Simulation 1 in Table \ref{tab1}, where the last column indicates the percentage of initial conditions resulting in horseshoe configurations. Green and black squares correspond to tadpoles and horseshoe orbits, respectively. We note that 26\% of the initial conditions result in mutual horseshoe orbital configurations between the minor satellites. Moreover, according to Fig. \ref{fig1}, the percentage of final horseshoe orbits decreases for smaller $(k_2/Q)_P$, indicating that a slow orbital decay scenario (implying in small dissipation within Saturn) is not favorable to explain the occurrence of mutual horseshoe regimes between Janus and Epimetheus.

\begin{figure}
\centering
%\subfigure{\includegraphics[width=0.43\textwidth]{mapa_2sig.eps}}
%\subfigure{\includegraphics[width=0.43\textwidth]{mapa_05sig.eps}}
%\subfigure{\includegraphics[width=0.43\textwidth]{fig1d.eps}}
\includegraphics[width=0.23\textwidth]{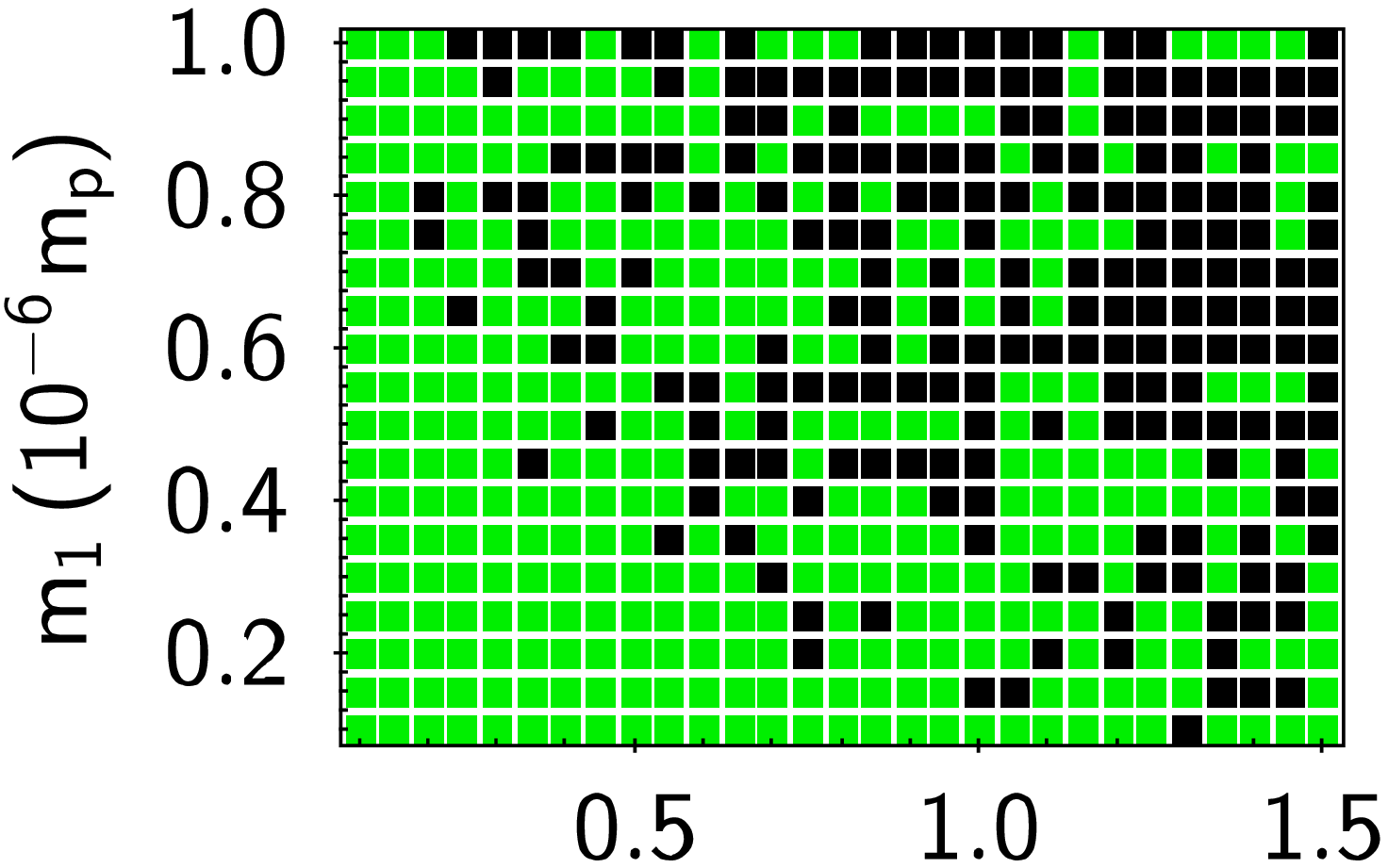}
\includegraphics[width=0.19\textwidth]{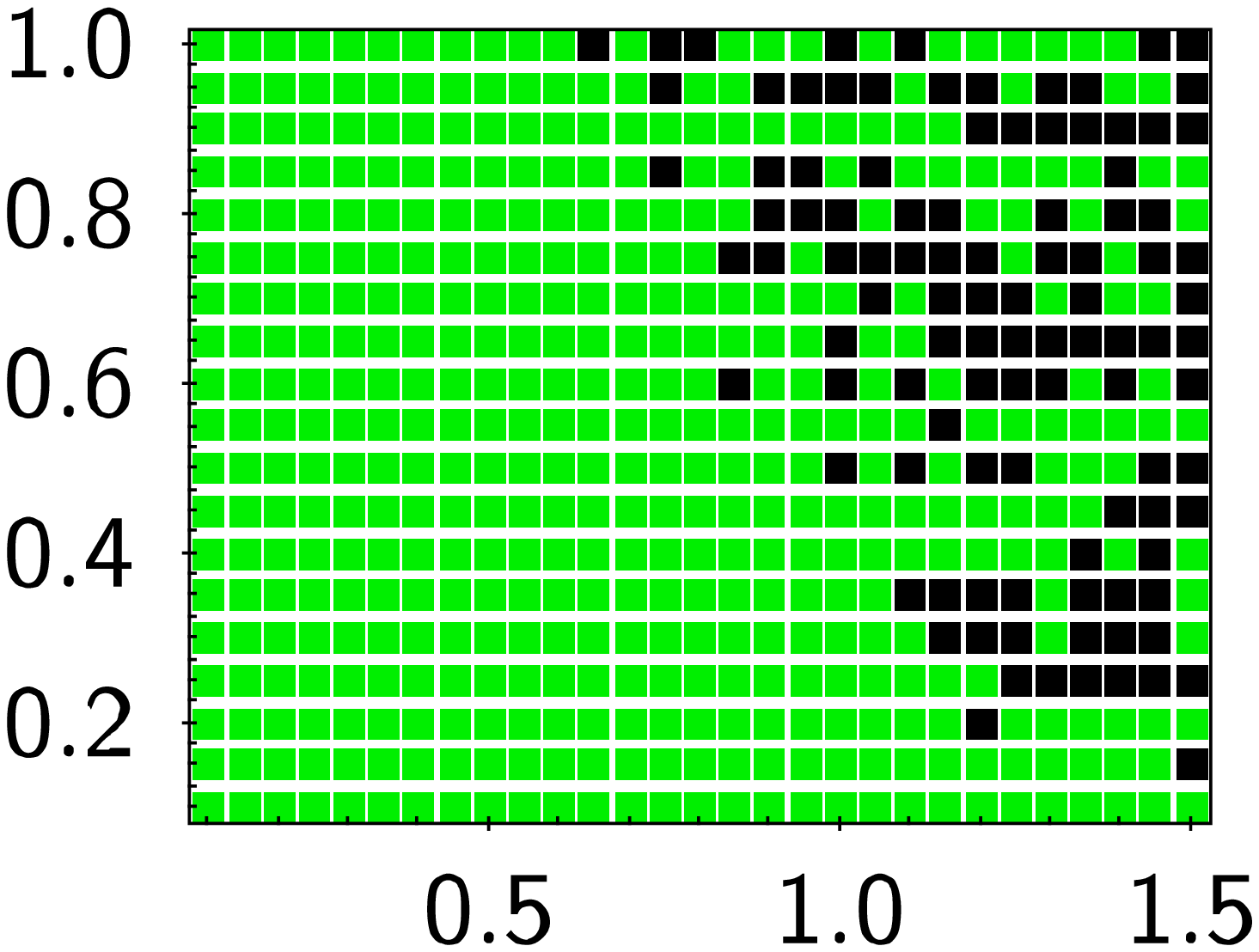}
\includegraphics[width=0.23\textwidth]{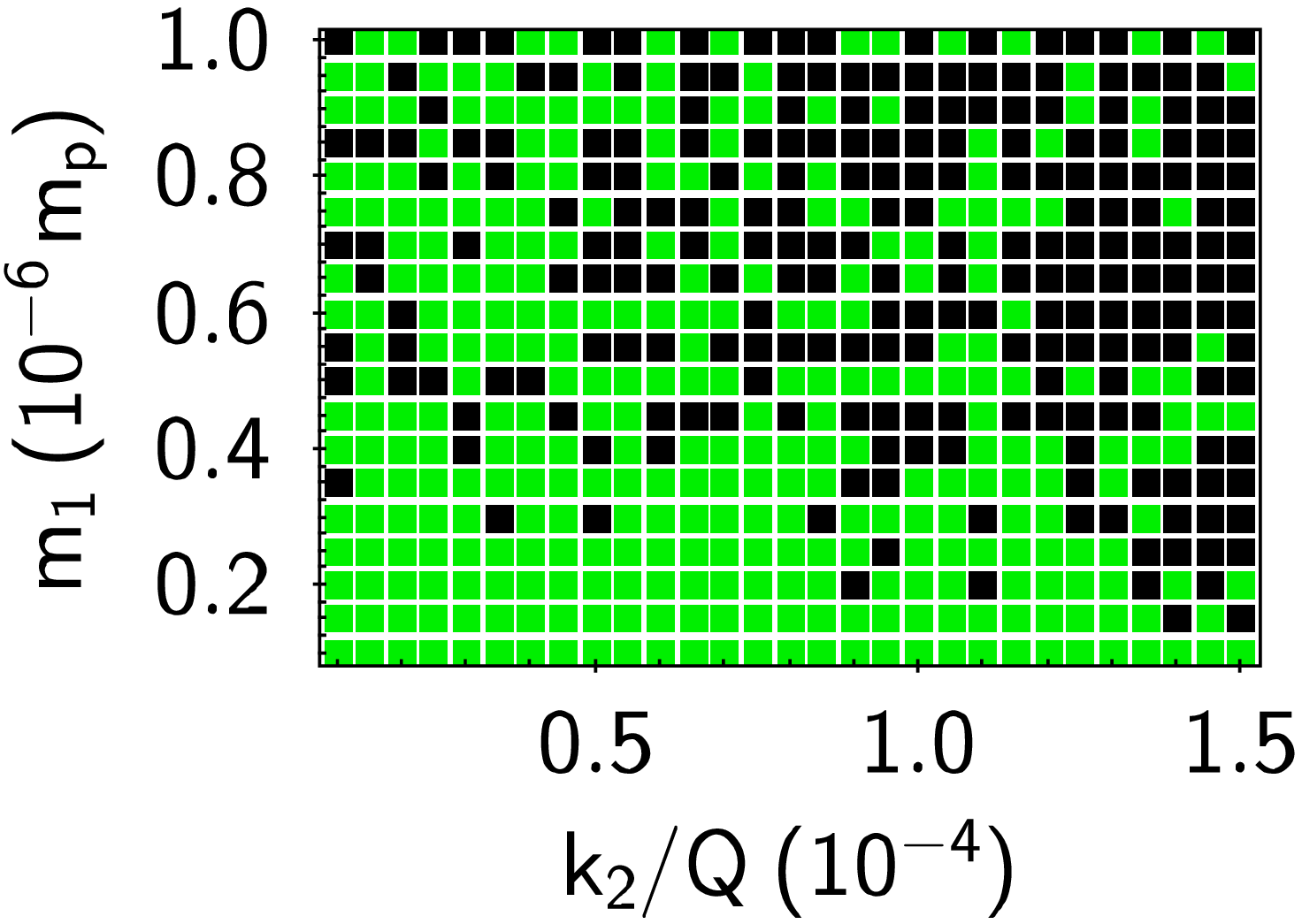}
\includegraphics[width=0.2\textwidth]{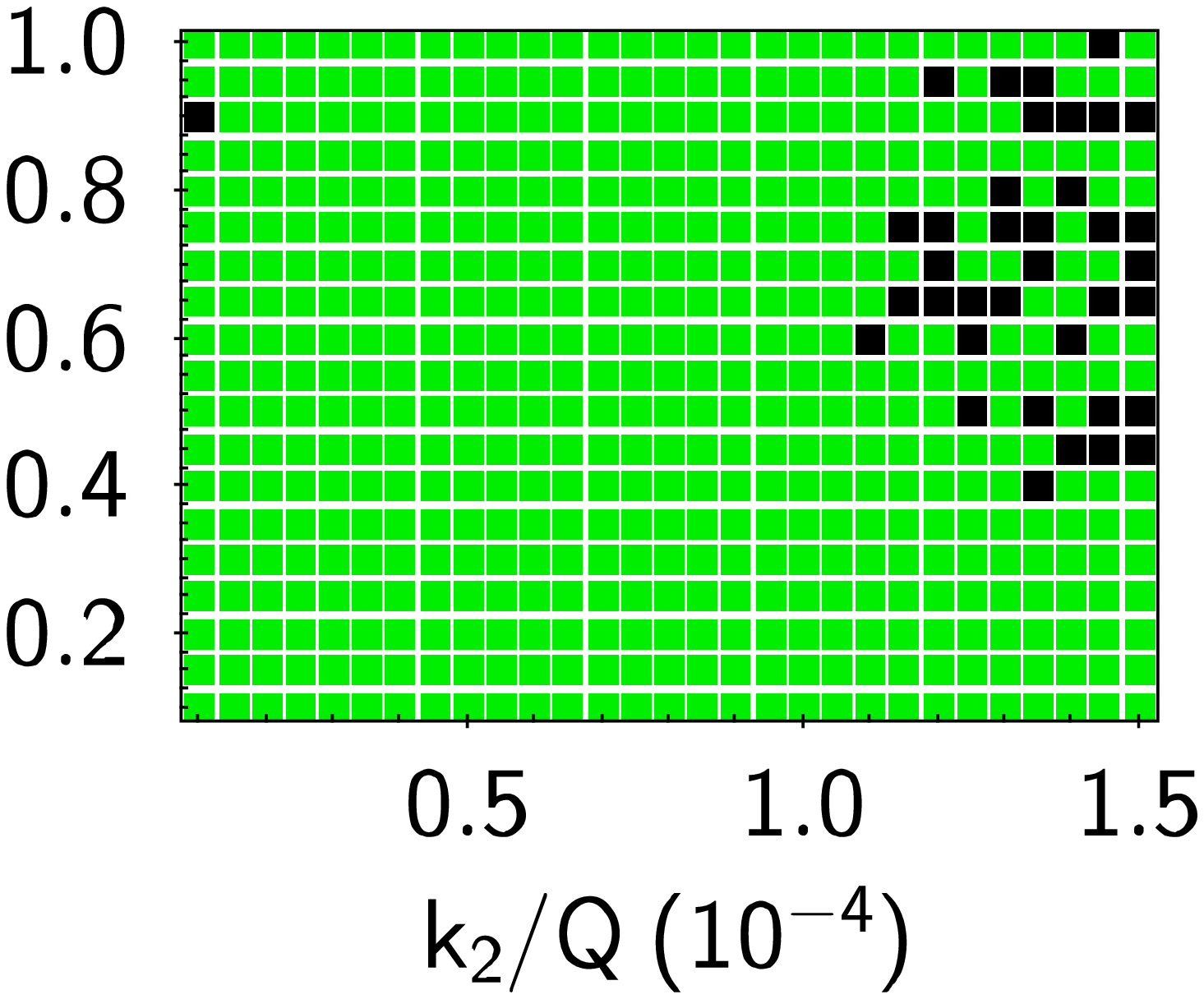}
\caption{{\small Numerical simulations for a grid in initial conditions as in Fig. \ref{fig1}, now changing the initial angular separation between the smallest satellites, $\Delta\lambda$, according to Table \ref{tab1}. From left to right and top to bottom, the panels correspond to the simulations 2 to 5 indicated in Table \ref{tab1}.}}
\label{fig3}
\end{figure}

In order to investigate whether the obtained results are affected by changing the initial angular configuration of the satellites, we perform additional sets of numerical simulations varying the initial mean longitudes, as shown in Table \ref{tab1} (Simulations 2 to 5). Fig. \ref{fig3} shows that when S2 is shifted from $L_4$, with S3 remaining in $L_5$ (Simulation 2, top left panel), the occurrence of horseshoe orbits is almost 1.5 times larger in comparison with the first set of simulations (compare Simulations 1 and 2 in Table \ref{tab1}). On the contrary, changes in the initial position of the minor satellite (S3) does not have significant implications for the final number of horseshoe orbits, as can be seen in Simulations 1 and 3 in Table \ref{tab1} and the top-right panel in Fig. \ref{fig3}. However, when both minor satellites are moved from $L_4$ and $L_5$ (Simulations 4 and 5) the results show high and low occurrence of final horseshoe orbits (bottom panels in Fig. \ref{fig3}). In summary, for initial $\lambda_2>60^{\circ}$ we obtain the largest number of final horseshoe co-orbital locking in our numerical simulations.

 \begin{figure}
\centering
%\subfigure{\includegraphics[width=0.43\textwidth]{mapa_2sig.eps}}
%\subfigure{\includegraphics[width=0.43\textwidth]{mapa_05sig.eps}}
%\subfigure{\includegraphics[width=0.43\textwidth]{fig1d.eps}}
\includegraphics[width=0.47\textwidth]{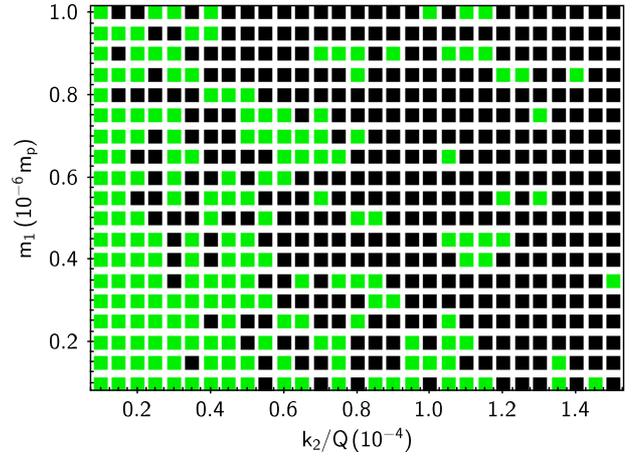}
\caption{{\small Numerical simulations for a grid of initial conditions as in Fig. \ref{fig1}, now changing the masses $m_2$ and $m_3$ according to Table \ref{tab1}. The figure corresponds to the simulation 6 indicated in Table \ref{tab1}.}}
\label{fig2}
\end{figure}

Looking at Figs. \ref{fig1} and \ref{fig3} we note that, for high $(k_2/Q)_P$ (high migration rates), the probability to found horseshoe orbits, as final orbital configuration between the minor satellites, is larger than for small $(k_2/Q)_P$ values. High value of $(k_2/Q)_P$ is in agreement with recent determinations for the dissipation within Saturn from astrometric measurements of their natural satellites (e.g., Lainey et al. 2017). On the other hand, the dependence on the mass of the guiding satellite ($m_1$) is not clear, although there is a slight trend to get more horseshoe orbits for large $m_1$.

We also performed additional sets of numerical simulations modifying the initial positions of the minor satellites, placing the largest one at (or close to, according to the same displacement indicated in Table \ref{tab1}) $L_5$ and the smaller one at (or close to) $L_4$. The results did not show significant differences with the above sets of numerical simulations shown in Table \ref{tab1}.

As mentioned in Sec. \ref{init}, we also change the masses of the minor satellites by an order of magnitude smaller and larger than the masses of Janus and Epimetheus (Simulations 6 and 7). We can see form Table \ref{tab1} and Fig. \ref{fig2} that the occurrence of horseshoe orbits increases for smaller masses of S2 and S3. Note that the value of $m_2/m_3$ does not vary in Simulations 1, 6 and 7, indicating that the final orbital configuration in horseshoe-like orbit depends on the individual masses of the minor satellites.

For sake of completeness, we performed additional simulations, now including companion satellites,  and analyzed their influence on the final configuration of the system under study. We included the satellites Prometheus and Pandora, which are located close to Janus and Epimetheus, but not Pan, Daphnis and Atlas due to their extremely small masses. We repeated the simulations shown in Fig. \ref{fig1}, showing that there is no significant modifications of the previous percentage of horseshoe and tadpoles final orbits.

\section{Discussion and conclusions}\label{conclusions}

In this paper, we analyze the orbital evolution of a system composed of a central planet and three co-orbital satellites. Our model accounts for gravitational interactions between the bodies, a circumplanetary gas disc and tidal interactions. We assume that two (less massive) satellites are initially located close the triangular equilibrium points ($L_4$ and $L_5$) with respect to the planet and the large (guiding) satellite (Fig. \ref{modelo}). Disc and tidal interactions result in orbital decay of the satellite system, in such a way that the three satellites remain in the co-orbital configuration, with the minor ones in a tadpole regime of motion. Through the solution of the exact equations of motions, we analyze the orbital behavior of the system under assumption that the guiding satellite is completely destroyed soon after it crosses the Roche limit with the planet. Several sets of numerical simulations are performed varying the the mass of the major satellite ($m_1$), the migration velocity (parameterized by $(k_2/Q)_P$) and the initial angular separation of the minor satellites with respect to the exact locations of $L_4$ and $L_5$ equilibrium points (defined by $\Delta\lambda$).

The results obtained show that, as soon as the guiding satellite crosses its Roche limit, either tadpole or horseshoe regimes of co-orbital motion appear as final configuration between the minor satellites. By varying $m_1$ and $(k_2/Q)_P$ , for a system with Saturn as the central planet and Janus and Epimetheus as the minor satellites, we found that 26\% of the numerical simulations result in horseshoe orbits for an initial angular configuration in which Janus and Epimetheus started at the $L_4$ and $L_5$ equilibrium points. Initial displacements of angular positions from $L_4$ or/and $L_5$ may significantly change the occurrence of the final horseshoe configurations (see Table \ref{tab1}). Moreover, decreasing or increasing the masses of the minor satellites by one order of magnitude results in drastic changes in the final orbital configurations, with a high occurrence of horseshoe orbits for smallest satellites. The migration rate also plays an important role, since a fast orbital decay favors the occurrence of horseshoe orbits in the numerical simulations.

The results obtained indicate that the currently observed horseshoe configuration between Janus and Epimetheus can be explained in the context of an early migration process interrupted by a cataclysmic event, in which their guiding co-orbital satellite is tidally disrupted via Roche limit crossing. Long-term numerical simulations should be carried out in order to investigate the stability of the obtained horseshoe orbits, however, this is out of the scope of the present paper and can be addressed in a further work.

%It is worth noting that, in this work, we do not intend to explain the exact current orbital configuration of Janus and Epimetheus. Indeed, for sake of simplicity, we do not considered some important effects, such as the interaction with neighbor planets or planetary rings. However, in the early Solar System, the orbits of the saturnian satellites may have affected by secular resonances with the giant planets (\'Cuk et al., 2018). Moreover, the interaction of Janus and Epimetheus with Saturn's rings could induce a transition from the current horseshoe-type lock to a tadpole orbit in a timescale of tens of millions years (Lissauer et al., 1985; Caudal, 2013). In order to get a complete analysis we would include such effects, however, the present work may help to give an insight into the observed horseshoe orbital confinement of these small saturnian  satellites. Moreover, it is not clear whether Janus and Epimetheus are recent (Charnoz et al. 2010) or primordial (Rossignoli et al., 2019) in their origin. Anyway, our model highlights the possibility to reproduce the mutual co-orbital configuration between km-sized satellites, in the horseshoe regime of motion, from the tidal disruption of a migrating co-orbital large satellite.

It is worth noting that, in this work, we do not intend to explain the current orbital configuration of Janus and Epimetheus. Indeed, for sake of simplicity, we do not considered some important effects, such as the interactions with neighbor planets or planetary rings. For instance, \'Cuk et al. (2018) have shown that secular resonances with the giant planets may have affected the orbital behaviour of the saturnian satellites in the early Solar System. Moreover, the interaction with Saturn's rings could induce a transition from the current horseshoe-type lock of Janus and Epimetheus to a tadpole orbit in a timescale of tens of millions years (Lissauer et al., 1985; Caudal, 2013). In addition, it is not yet clear whether Janus and Epimetheus are recent (Charnoz et al. 2010) or primordial (Rossignoli et al., 2019) in their origin. However, although our study does not consider such effects, it points out the possibility to reproduce the mutual co-orbital configuration between km-sized satellites, in the horseshoe regime of motion, through the tidal disruption of a migrating co-orbital large satellite. Thus, we expect that the present work gives an insight into the observed horseshoe orbital confinement of the small saturnian satellites. 

\vspace{3cm}

%- comparar algum caso com oscilacao atual de JE?

%\section{Bibliography}

%\begin{itemize}
%\item Canup R. M. \& Ward W. R. A., 2006, Nature, Volume 441, pp. 834-839
%\item Canup R. M., 2010, Nature,  Volume 468, Issue 7326, pp. 943-946
%\item Mosqueira I., and Estrada P. R., Icarus, 2003, 163, 198
%\item Tanaka, H., \& Ward, W. R. 2004, ApJ, 602, 388
%\item Tanaka, H., Takeuchi, T., \& Ward, W. R. 2002, ApJ, 565, 1257
%\end{itemize}

\subsection*{Acknowledgements}
We acknowledge our referee, Dr. Maryame El Moutamid, for her detailed review and for the helpful suggestions, which allowed us to improve the manuscript. This work was supported by the Brazilian CNPq, CAPES and FAPESP agencies. J. Correa-Otto gratefully acknowledges the financial support by CONICET through PIP 112-201501-00525, and the partial financial support by CICITCA, UNSJ through the projects 21/E1079 (2018-2019) and PROJOVI (2018-2019).

\end{document}